\documentstyle[twoside,epsf,fleqn,espcrc2]{article}

\newcommand{\AmS}{{\protect\the\textfont2
  A\kern-.1667em\lower.5ex\hbox{M}\kern-.125emS}}

\title{
\vspace*{-15pt}
{\normalsize November 1992 \hfill UTHEP-246}
Contamination~of~Excited~States~in~Quenched~QCD~Hadron~Propagators\thanks{
Talk presented by T.Yoshi\'e at Lattice '92}}

\author{QCDPAX collaboration: \\
Y.~Iwasaki
\address{Institute of Physics, University of Tsukuba, Ibaraki 305, Japan},
K.~Kanaya$\mbox{}^{\rm \ a}$,
S.~Sakai
\address{Faculty of Education, Yamagata University, Yamagata 990, Japan},
T.~Yoshi\'e$\mbox{}^{\rm \ a}$,
T.~Hoshino
\address{Institute of Engineering Mechanics, University of Tsukuba,
Ibaraki 305, Japan},
T.~Shirakawa$\mbox{}^{\rm \ c}$
and
Y.~Oyanagi
\address{Department of Information Science, University of Tokyo,
Tokyo 113, Japan}
}

\begin{document}

\begin{abstract}
Quenched QCD hadron spectrum is calculated with Wilson's quark action at
$\beta=5.85$ and 6.0 on a $24^3 \times 54$ lattice.
We discuss the problem of whether
we have extracted the mass of the ground state at these $\beta$'s
without contamination of the excited states.
We show that the masses of the first excited states turn out to be consistent
with experiment
when we are able to obtain the propagators up to the large time slice region
where the propagators take the asymptotic
forms of the ground states.
\end{abstract}

\maketitle

Deriving the hadron spectrum from first principles of QCD has been
a long standing challenge for lattice QCD simulations.
As a step toward this goal, the QCDPAX Collaboration has
carried out a high statistics quenched calculation
with Wilson quarks at $\beta$= 5.85 and 6.0 on a $24^3 \times 54$ lattice.
After we reported the results of a preliminary analysis
at Lattice 91\cite{Lat91},
we have increased statistics at $\beta$=6.0.
The results reported here are based on an analysis of 100(200) configurations
at $\beta$=5.85(6.0),
each configuration being separated by 1000 heat bath sweeps.
Hopping parameters are chosen to be $K$=0.144,0.154,0.1585,0.1595 and
0.1605 at $\beta$=5.85 and $K$=0.145,0.152,0.155,0.1555 and 0.1563 at
$\beta$=6.0, respectively.
They are chosen in such a way that $m_{\pi}/m_{\rho}$ roughly agree
with each other. The ratio ranges from 0.97 to 0.53.
The errors we quote are estimated by the jack-knife method.

In this article, we discuss the problem of whether the hadron propagators
calculated at these $\beta$'s have
really taken the asymptotic forms of the ground states at large
time slices.
Recently, systematic errors due to contamination of the excited states
at $\beta \sim$ 6.0 for Wilson quarks have been much reduced by
calculations on lattices with large size
in time direction\cite{Lat91,Lat89} and
calculations using smeared quark sources\cite{Ape60,Hemcgc,Lanl,Gf11}.
(See also refs.\cite{Ape63,Ukqcd,Qcdtaro}
for works at larger $\beta$'s.)
However, we think that the problem has not been settled yet.
For example, Los Alamos group\cite{Lanl} has obtained different mass
results from different types of quark sources.
Ape group\cite{Ape60,Ape63} has found that the masses of the first
excited states of $\pi$, $\rho$ and nucleon come out much heavier
than experiment.
Therefore, we have made various analyses, using our data which are obtained
with point quark source, to see whether the ground states dominate the hadron
propagators at large $t$'s
and whether the masses of the first excited states turn out to be
reasonable.

\begin{figure}[t]
\begin{center}
\leavevmode
\epsfysize=200pt
  \epsfbox{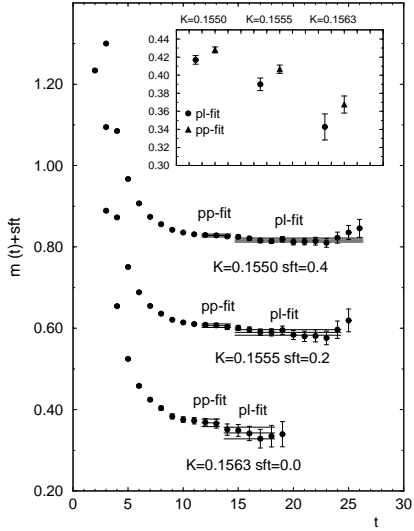}
\end{center}
\vskip -5mm
\caption{Effective mass plots for $\rho$ meson at $\beta=6.0$ for the
largest three hopping parameters.
In the small box shown are masses obtained by two different fits.}
\label{rho-b600}
\end{figure}

First, we discuss the results for $\rho$ meson.
We find that there exists a plateau in effective mass $m(t)$
in the sense that the $m(t)$ does not vary more than one
standard deviation for successive values of time slices.
In fig.~\ref{rho-b600} we show the results for the $m(t)$ at $\beta$=6.0.
The results at $\beta$=5.85 are similar.
The $m(t)$'s reach a plateau at $t_1 \sim 11$ at $\beta$=5.85 and typically at
$t_1\sim 15$ at $\beta$=6.0, respectively.
(Note that the $\beta$ dependence of the time slice where the plateau
starts at is consistent with the ratio of the lattice spacings
$a^{-1}(\beta=6.0)/a^{-1}(\beta=5.85) \sim 1.25$.
We have estimated the ratio from the $m_{\rho}$ at $K_c$.)
Therefore we have fitted propagators for $t \ge t_1$ to single
hyperbolic cosine functions.
Hereafter we call this fit plateau-fit or pl-fit.
The $m(t)$'s do not reach a plateau at $t<t_1$.
To see this, we have made improper fits with data for several $t$'s
with $t<t_1$.
Hereafter we call this fit pre-plateau fit or pp-fit.
The mass results obtained by the pre-plateau fits are about two standard
deviations larger than those by the plateau fits.
(See fig.~\ref{rho-b600}.)
At $\beta$=6.0, the differences of the two results are about 3\% at $K$=0.155
where $m_{\pi}/m_{\rho} \sim$ 0.7,
and about 7\% at $K$=0.1563 where the ratio is about 0.5.

\begin{figure}[htbp]
\begin{center}
\leavevmode
  \epsfysize=200pt
  \epsfbox{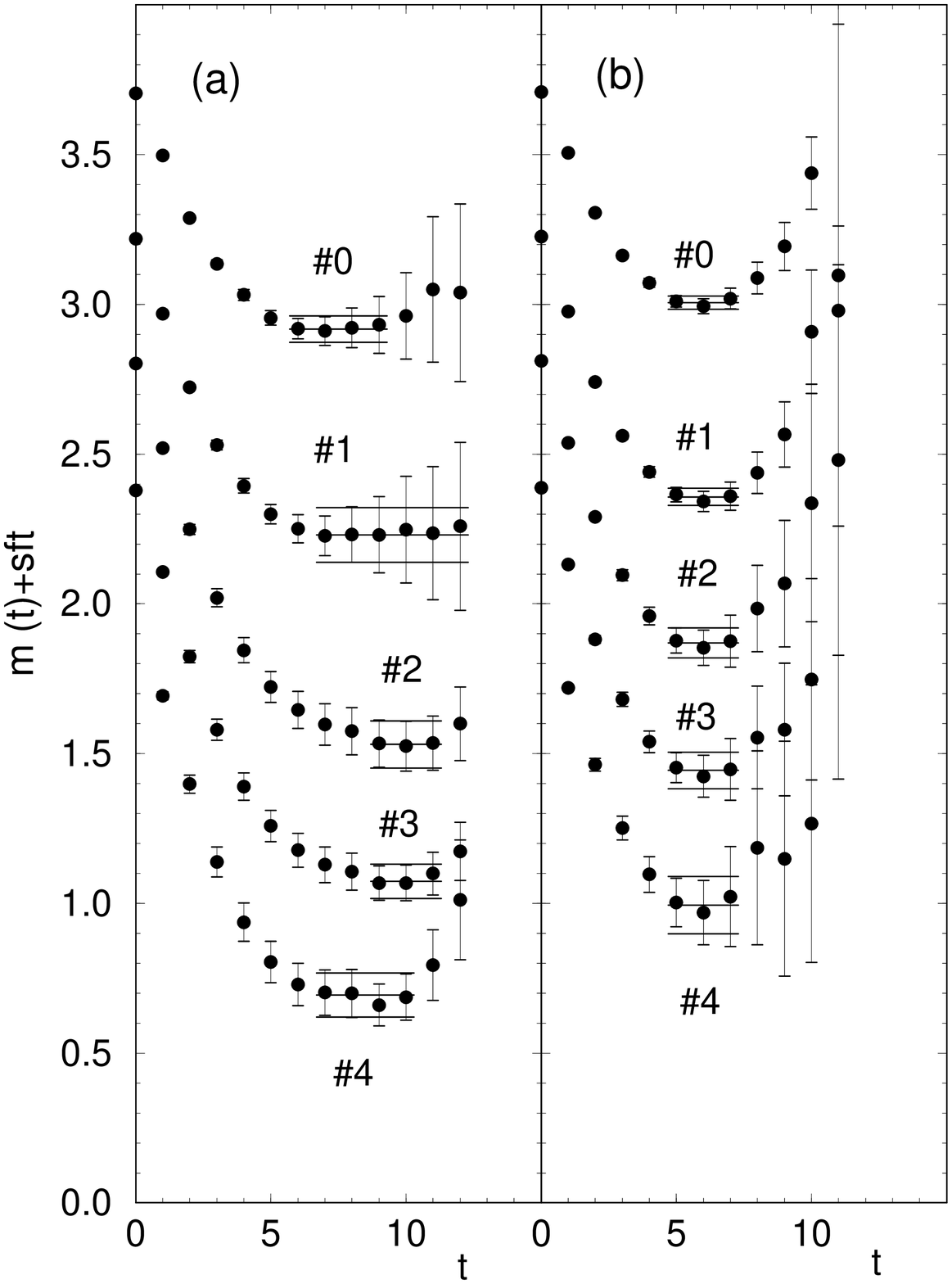}
\end{center}
\vskip -5mm
\caption{Effective mass plots for the first excited state of $\rho$ meson
at $\beta$=6.0. (a) the case where the plateau fits are used for
the ground state. (b) the case where the pre-plateau fits are used.
\#'s denote hopping parameters:
\#0:$K$=0.145(sft=1.6),\ \#1:$K$=0.152(sft=1.2),
\#2:$K$=0.155(sft=0.8),\ \#3:$K$=0.1555(sft=0.4),
\#4:$K$=0.1563(sft=0.0).}
\label{rho-ex-b600}
\end{figure}

\begin{figure}[htbp]
\begin{center}
\leavevmode
  \epsfysize=150pt
  \epsfbox{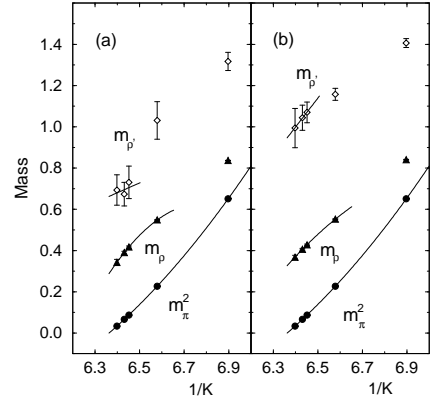}
\end{center}
\vskip -5mm
\caption{Mass vs. $1/K$ for $\rho$ and $\rho'$ at $\beta$=6.0.
(a) the case where the plateau fits are used for the ground state.
(b) the case where the pre-plateau fits are used.}
\label{kinv-rho-b600}
\end{figure}

The two different fits for the ground state lead a drastic difference
in the results for the mass of the first excited state $\rho'$.
In order to estimate $m_{\rho'}$, we calculate excited
state propagators defined by
\begin{equation}
G'(t)=G(t) - A_0 {\rm cosh}((T-t/2)m_0)
\end{equation}
with $A_0$ and $m_0$ determined by the pl-fits or the pp-fits.
Fig.~\ref{rho-ex-b600} shows effective mass plots for the $G'(t)$.
When the plateau fit is used for the ground state,
there exists a plateau in effective mass $m(t)$ for the $G'(t)$.
When the pre-plateau fit is used, no clear plateau can be seen
in the $m(t)$ for the $G'(t)$.
The $m_{\rho'}$ is determined by fitting the $G'(t)$ for time slices shown in
fig.~\ref{rho-ex-b600}.
Fig.~\ref{kinv-rho-b600} shows the masses thus obtained
versus $1/K$ at $\beta$=6.0.
The results at $\beta$=5.85 are similar.
When the plateau fit is used for the ground state,
the $m_{\rho'}$ turns out to be about twice the $m_{\rho}$:
the $m_{\rho'}$ at $K_c$ in physical units read
1.57(21) GeV at $\beta$=5.85 and 1.77(24) GeV at $\beta$=6.0, respectively,
which are consistent with experimental value 1.6 GeV.
When the pre-plateau fit is used, the $m_{\rho'}$ is about three times
the $m_{\rho}$, or in physical units 2.30(17) GeV at $\beta$=5.85 and
2.23(24) GeV at $\beta$=6.0, respectively.
These values are much larger than the experimental value.

If lattice QCD at these $\beta$'s in the quenched approximation
has something to do with continuum QCD,
we expect that we can obtain a reasonable value for the mass of the first
excited state as well as that for the mass of the ground state.
It may be worth while emphasizing that
the mass of the first excited state turns
out to be consistent with experiment,
when we fit data for time slices
where the effective mass plot has a plateau for the ground state.

\begin{figure}[t]
\begin{center}
\leavevmode
  \epsfysize=150pt
  \epsfbox{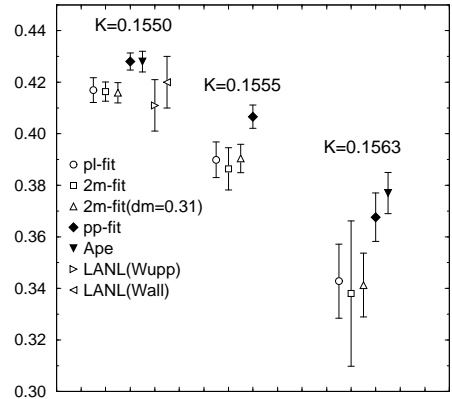}
\end{center}
\vskip -5mm
\caption{Mass results for $\rho$ meson at $\beta$=6.0 obtained by various fits
as well as those reported by Ape group\protect\cite{Ape60} \protect
and Los Alamos group\protect\cite{Lanl} \protect.}
\label{rho-mass-b600}
\end{figure}

We have also made two-mass fits and two-mass fits with mass difference fixed
to its experimental value.
Fig.~\ref{rho-mass-b600} summarizes the results
for the $m_{\rho}$ at $\beta$=6.0
obtained by various fits as well as those reported by Ape group
\cite{Ape60} and Los Alamos group\cite{Lanl}.
The results obtained by the plateau fits, the two-mass fits and the two-mass
fits with mass difference fixed are consistent with each other.
The results by the pre-plateau fits and those reported by Ape group are
consistent with each other.
However the latter is about two standard deviations larger than the former.
This deviation causes a large difference for the mass of the first excited
states.

\begin{figure}[htbp]
\begin{center}
\leavevmode
  \epsfysize=150pt
  \epsfbox{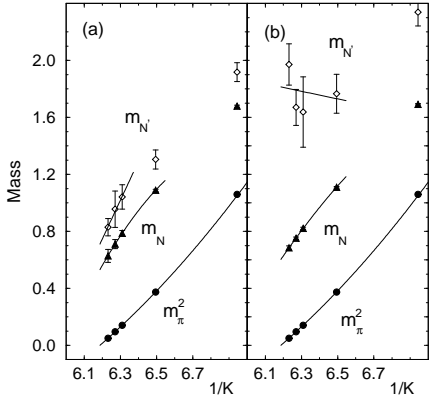}
\end{center}
\vskip -5mm
\caption{The same as fig.~\protect\ref{kinv-rho-b600} \protect
but for nucleon at $\beta$=5.85.}
\label{kinv-proton-b585}
\end{figure}

We have made similar analyses for nucleon at $\beta$=5.85.
Effective masses for the ground state reach a plateau typically at $t \sim 15$
for small quark masses, except at $K=0.1605$ where we
have not obtained clear plateau.
We find that there exists a plateau in effective masses
also for the first excited state $N'$,
when the plateau fits are used for the ground state.
The $m_{N'}$ turns out to be 1.53(22) GeV,
which is consistent with experimental value 1.44 GeV.
(See fig.~\ref{kinv-proton-b585}.)

\begin{figure}[t]
\begin{center}
\leavevmode
  \epsfysize=200pt
  \epsfbox{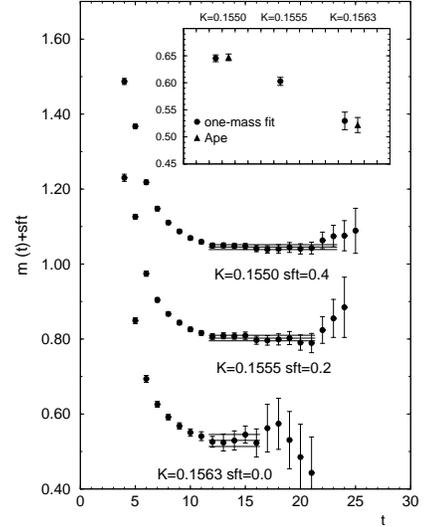}
\end{center}
\vskip -5mm
\caption{The same as fig.~\protect\ref{rho-b600} \protect but for Nucleon at
$\beta$=6.0.}
\label{pro-b600}
\end{figure}

For nucleon at $\beta$=6.0, we have first fitted data for $t \ge 12$,
because at first sight, it seems that there exists a plateau in
effective masses for $t \ge 12$. (See fig.~\ref{pro-b600}.)
The results for $m_{N}$ obtained by the one-mass fits
are consistent with those reported by Ape group.
However, the mass of the first excited state $m_{N'}$ turns out to be
much heavier than experiment: $m_{N'}$ at $K_c$ is 2.91(58) GeV.
(This value is also consistent with that of Ape group.)

We interpret this due to the fact
that contamination from the excited states
is still large around $t \sim 12$,
because $t \sim 12$ at $\beta$=6.0 corresponds to $t \sim 10$ at $\beta$=5.85,
where we have found the plateau in effective masses for the first excited
state of nucleon at $\beta$=5.85.
When
we recall that the ratio of the lattice spacings at two $\beta$'s is about 1.25
and the $m(t)$'s for nucleon at $\beta$=5.85 reach a plateau at $t \sim 15$,
we expect that the $m(t)$'s at $\beta$=6.0 reach a plateau at $t \sim 19$.
Our data for the effective masses at such large time slices are not good enough
to confirm this expectation.
However we find that the mass results obtained by two-mass fits with
mass difference fixed are slightly smaller than those by the one-mass fits.
This suggests that the nucleon masses at $\beta$=6.0 can be smaller
that the results obtained by the one-mass fits.

The other extreme explanation is that
the masses obtained by the one-mass fits
as well as those reported by Ape group are correct,
and therefore, the mass of the first excited state is really heavy.
In this case, $N'$(1.44 GeV), which is well established by experiment,
should be an exotic state.

{}From analyses above, we interpret our results as follows.
Propagators of $\rho$ meson at $\beta$=5.85 and 6.0 as well as those of
nucleon at $\beta$=5.85 have taken at large $t$'s
the asymptotic forms of the ground states.
In these cases, masses of the first excited states turn out to be reasonable.
For nucleon at $\beta$=6.0, the first exited state turns out to be much heavier
than experiment, when the one-mass fits for $t \ge 12$ are used
for the ground state.
In order to obtain definitive results, we need good data for $t \ge 19$.
We hope we can perform high statistics calculations on larger lattices
in time direction, or calculations with some kind of smeared quark sources
at $\beta$=6.0, in order to obtain data for larger time slices.

The numerical calculations have been performed with QCDPAX,
a parallel computer developed at University of Tsukuba.
We would like to thank Rajan Gupta for valuable discussions.
This project is supported by the Grant-in-Aid
of Ministry of Education, Science and Culture
(No.62060001 and No.02402003).

\end{document}